\documentclass[12pt]{iopart}

\usepackage{iopams}
\usepackage{graphicx}
\begin{document}

\letter{Walkers on the circle}

\author{Daniel Jezbera$^1$, David Kordek$^1$, Jan K\v r\'\i \v z$^{1,3}$,
Petr \v Seba${}^{1,2,3}$ and Petr \v Sroll$^1$}

\address{$^1$ University of Hradec Kr\'alov\'e, Rokitansk\'eho 62, Hradec Kr\'alov\'e,
Czech Republic}
\address{$^2$ Institute of Physics, Academy of Sciences of the
Czech Republic, Na~Slovance~2, Prague,  Czech Republic}
\address{$^3$ Doppler Institute for Mathematical Physics and Applied
Mathematics, Faculty of~Nuclear Sciences and Physical Engineering,
Czech Technical University, B\v rehov\'a~7, Prague, Czech Republic}
\ead{\mailto{daniel.jezbera@uhk.cz},\mailto{david.kordek@uhk.cz}, \mailto{jan.kriz@uhk.cz},
\mailto{seba@fzu.cz}, \mailto{petr.sroll@uhk.cz}}
\begin{abstract}
We experimentally demonstrate that the statistical properties of
distances between pedestrians which are hindered from avoiding each
other are described by the Gaussian Unitary Ensemble of random
matrices.
The same result has recently been obtained for an $n$-tuple of non-intersecting
(one-dimensional, unidirectional) random walks.
Thus, the observed behavior of autonomous walkers conditioned not to cross their
trajectories (or, in other words, to stay in strict order at any time)
resembles non-intersecting random walks.
\end{abstract}

\vspace{2pc}
\maketitle

The fact that non-intersecting one-dimensional random walks lead to
universal system behavior has been known and discussed for at least 10
years \cite{hob}, \cite{grab}, \cite{kon}. It is also known that the
results can be described in terms of random matrix theory - see
for instance \cite{joh}. This fact is usually expressed in abstract
mathematical theorems of universal validity, see for instance~\cite{eichel}.

Our aim here is to use these abstract results in order to explain certain
aspects of the observed behavior of pedestrians. A comprehensible application of
the complicated mathematical theory is given e.~g.~in~\cite{baik}. It
analytically explains an experimental observation that the schedule of the
city transport in Cuernavaca (Mexico) conforms to the predictions of
the Gaussian Unitary Ensemble of random matrices (GUE). The reason
for this interesting observation is the absence of a bus timetable,
and primarily the fact that the buses do not overtake each other and
hence their trajectories do not cross - see also \cite{seba} for the
details. Our focus in this letter is pedestrians in a situation
when they cannot avoid each other.

Pedestrian flow is a subject of intense study. This is
understandable since the movement of large groups of people inevitably leads
to injuries and deaths caused by trampling and by crowd-pressure.
The consequences can be disastrous: for instance more than
1400 people were trampled to death during a stampede in Mecca in 
1990. A proper understanding of the process how groups of people move is vital
for taking effective precautions. The mathematical description is usually based
on the pedestrian interactions denoted as "social forces" \cite{h1}. The exact
character of these forces and of their cultural dependence remains
unclear, however, and is a focus of recent discussions \cite{sey2}. An
evacuation dynamics of buildings is modeled in a similar way \cite{h2}. But not
only panic situations are of importance. The comfortable and safe movement of
people through corridors and on sidewalks is also of interest.  Although people
are autonomous individuals following their own destinations, they cannot move
freely as soon as the pedestrian density exceeds a certain limit. For higher
densities self-organizing phenomena occur. A typical example is the
stratification of pavement walkers into layers for different direction \cite{h4}.

We will discuss a unidirectional pedestrian motion in the range of
intermediate density and in a narrow corridor that hinders mutual
avoidance. Otherwise the people can move freely. Since to avoid
another walker is not possible, the walker's attention naturally focuses on the
preceding fellow, in order not to collide with him. The situation resembles the
assumptions of the model of vicious random walkers introduced by
Fisher \cite{fis}. In a typical case, the vicious walkers move
randomly on a one-dimensional discrete lattice. At each time step,
that walker can move either to the left or to the right. The only
constraint is that two walkers cannot occupy the same site at the
same time. The model is easily modified to the situation when the
motion is unidirectional (for instance right moving). In this case,
the walkers are staying at the same site instead of moving to the left. The
model has surprising relations with various fields of mathematics
like combinatorics or random matrix theory \cite{baik2}. The
corresponding random matrix ensemble is, however, not fixed solely by
the dynamics. It depends also on the particular initial and
terminal conditions of the model \cite{kat}.

Our measurement is inspired by the paper \cite{sey} where the
fundamental diagram (i.e. the dependence of the pedestrian flow on
the pedestrian density)  has been measured experimentally with
volunteers walking in a circle. Fundamental diagrams are one of the
basic tools used in car flow modeling and traffic jam
prediction. In highway traffic, its shape is influenced by many
factors like the road topology, the existence of a near slip road,
and so on. The interesting question of how cultural differences
influence the flow-density relation for pedestrians
 has been discussed in \cite{chat}. Beside the fundamental diagram,
the work also discusses density fluctuations which are of vital
interest for the stampede dynamics. In a crowd, it is sudden
density changes that lead to the abrupt release of the local pressure
and finally cause people to fall and be trampled \cite{h3}. In
a one-dimensional system, the local density is inversely
proportional to the local distance among the walkers (the pedestrian
clearance). So we will  discuss the statistical properties
of the experimentally measured pedestrian distances.

\begin{figure}
\begin{center}
  \includegraphics[width = 0.9\textwidth]{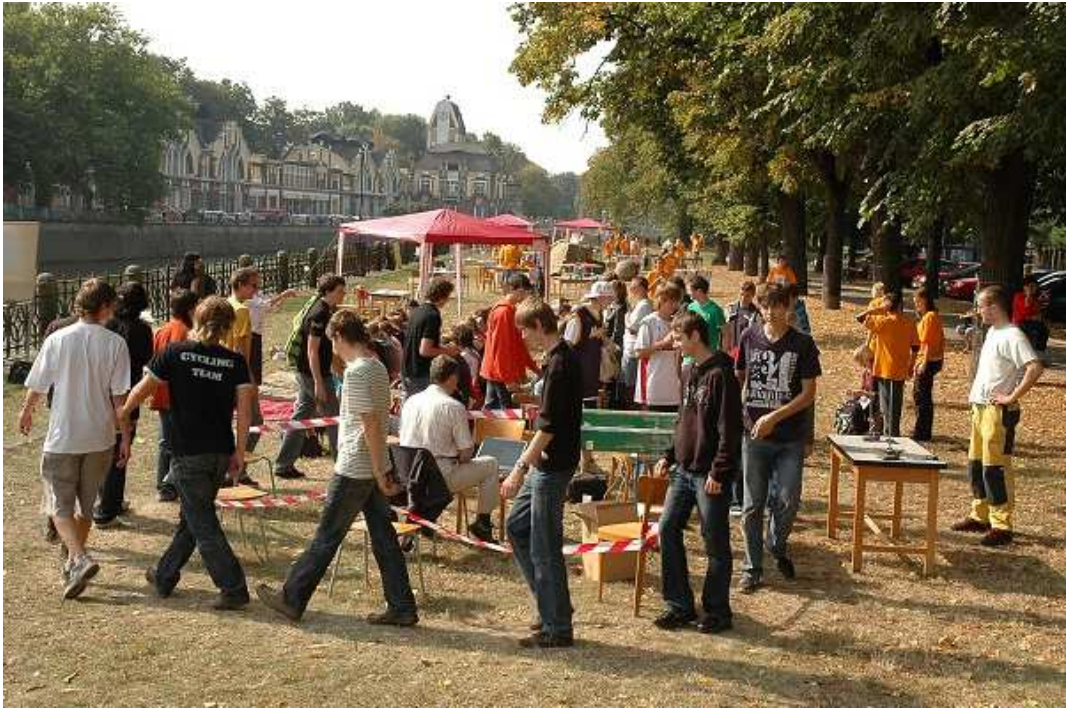}\\
\end{center}
  \caption{
  The measurement.
  }
  \label{fotka}
\end{figure}

During the public action called "Let us use our heads to play" (an event
serving to popularize physics among schoolchildren), we prepared a circular
corridor of a diameter of~4.5~m, built with chairs and ropes, see Figure~\ref{fotka}.
It was placed on a grass plot in front of the university building.
As various school classes participated in this activity, we asked
them  to walk in the corridor for a time
period of 3 minutes.
The motion of walkers was registered by two light gates
placed in a fixed distance of less than 1~m. The times when the light of the gates
was interrupted by the walker were recorded by a computer at a rate of 50
samples per second.
It means that at the mean walking speed of 1.25 m/s, the device resolution
was approximately 0.025~m.
The interruption onset of the gate was taken as the
pedestrian arrival.
In this way, we obtained the record of 26~groups
consisting of 12 to 20 walkers.
Using these data, it is easy to obtain the pedestrian headway
and velocity, and combining these quantities, the pedestrian clearance is obtained
straightforwardly.
We will, however, focus directly on the headway statistics, i.e.
the statistics of the time intervals between two subsequent walkers. To avoid
global density effects (like small jams inside the circle when one walker slows
down unexpectedly) the headway data of each particular measurement were
unfolded and scaled to a mean headway equal to one. The theory published in
\cite{baik} predicts that the headway of autonomous (random) pedestrians
hindered from avoiding each other is described by the level spacing
distribution of the Gaussian unitary ensemble of random matrices. In other
words, the headway statistics are universal and given by the Wigner formula for
GUE:
$$
    P(s)=\frac{32}{\pi^2} s^2 \exp\left(-\frac{4 s^2}{\pi}\right)
$$

\begin{figure}
\begin{center}
  \includegraphics[height=9cm,width=15cm]{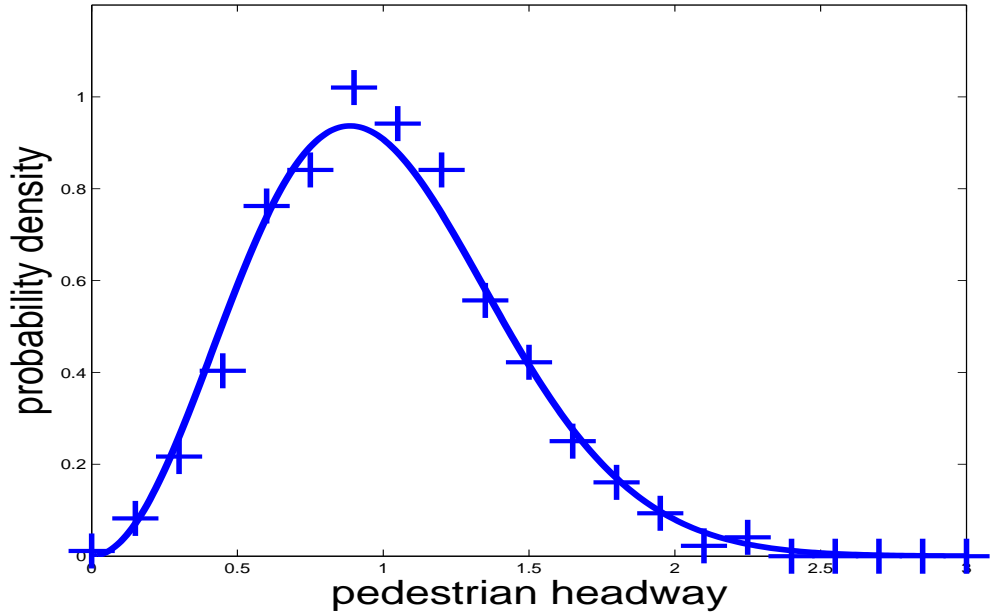}\\
\end{center}
  \caption{
  The headway probability density evaluated with data measured for 26 different pedestrian groups
  (crosses) is compared with the prediction of GUE (full line).
  The mean headway is scaled to unity.
  }
  \label{head}
\end{figure}

The results obtained for the circular corridor with 13-15 years old children
from 26 different groups is plotted in the Figure~\ref{head}.

As already mentioned, understanding the local fluctuations of the
density is of particular interest since the unexpected density
changes are suspected to be the main cause for people falling and being
trampled~\cite{h3}. In the one-dimensional pedestrian flow, the density
fluctuations were investigated in \cite{sey}, \cite{chat}. The
authors argued conjecturally that the fluctuations display a small dependence
on the cultural backgrounds of the walkers (Indian and German
pedestrians were compared). On the other hand, the mathematical
theory developed for non-crossing trajectories predicts a universal
behavior pattern. This means that the cultural factors should be statistically
irrelevant. The question is  how these two points can be reconciled. To
investigate it, we will study the number of people passing a given
point within certain time interval. Let $n(T)$ be the number of
walkers passing the measuring point within the time interval $T$.
The fluctuation $\Sigma(T)$ is defined as
$$
\nonumber \Sigma(T)=\left \langle \left( n(T)-\langle n(T)\rangle\right) ^2
\right \rangle $$
where $\langle ... \rangle$ means the system average. Traditionally,
this quantity is called the number variance, and its behavior is well
understood. For uncorrelated events (Poisson process) $\sigma(T)=T$
for large $T$. For the headway governed by the GUE ensemble, we get
$\Sigma(T)\approx (ln(2\pi T)+1.5772..)/\pi^2$ and the fluctuation
of the number of walkers passing a given point increases only
logarithmically with the time. Generally, point processes leading to
$\Sigma(T)<T$ for large $T$ are denoted as superhomogeneous (see for
instance \cite{tor1}, \cite{tor2}).

\begin{figure}
\begin{center}
  \includegraphics[height=9cm,width=15cm]{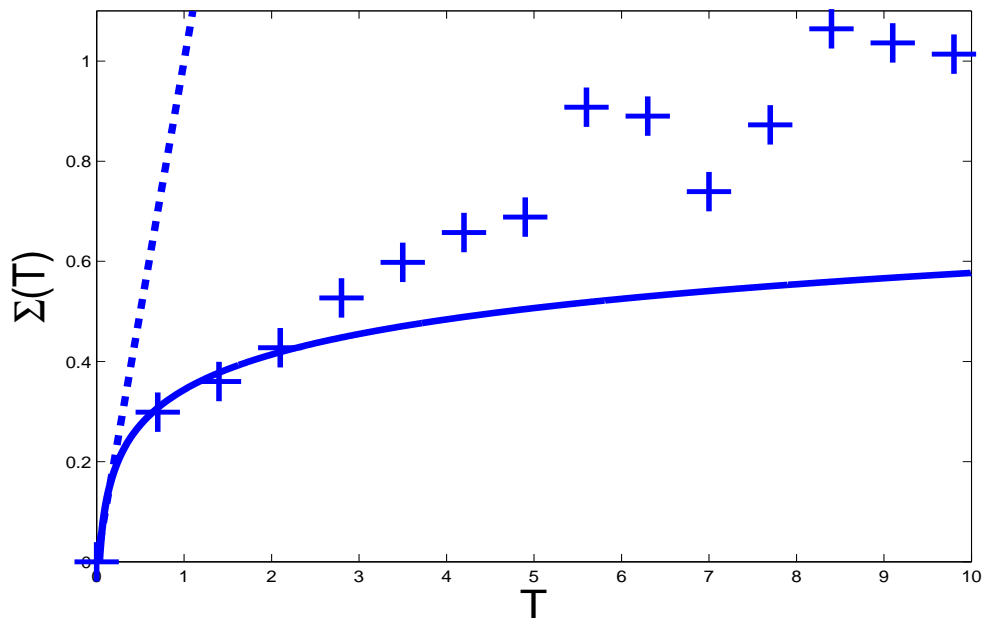}\\
\end{center}
  \caption{
   The number variance $\Sigma(T)$ evaluated with data measured for 26 different pedestrian groups
  (crosses) is compared with the prediction of GUE (full line) and with the prediction for a random headway sequence (dashed line).
  }
  \label{numb}
\end{figure}

We used the measured data to evaluate the number variance for the pedestrians
inside the circular corridor with foregoing restrictions. The results are plotted
in Figure~\ref{numb}. We see that $\Sigma$ follows the prediction of GUE up
to $T\approx 3$. For larger $T$ the increase is higher than the logarithmic
prediction of GUE. It remains, however, substantially below the line
$\Sigma(T)=T$ obtained for the uncorrelated events. So the headway sequence is
superhomogeneous and follows the GUE result for  $T\lesssim 3$. The cultural
differences reported in \cite{chat} can be related with the non-universal
increase of $\Sigma$ for $T \gtrsim 3$. Similar behavior has been also observed
for bus transport \cite{seba}.

The interactions between pedestrians are obviously asymmetrical -
the walker is observing the walker ahead of him (so as not to collide with him)
and has much less information on the walker behind him.
Thus the Newton's law of ``actio = reaction'' does not hold for pedestrians.
According to our results, the distance distribution, however, agrees with
classical many-particle systems,
such as Dyson's gas, where the interactions are symmetrical.
The statistical properties of one-dimensional many-particle systems violating
the ``actio=reactio'' law were studied in~\cite{TreiberHelbing}.
They have analytically shown that such systems exhibit the same statistics
as classical Newtonian systems.

To summarize: we have presented experimental data on pedestrians
walking in a circular corridor when they are hindered from avoiding each other. The
results are in agreement with the prediction obtained for the mathematical
models of one-dimensional vicious random walkers.
Our results can also be regarded as (another) experimental demonstration of the
theoretical results of~\cite{TreiberHelbing}.
\ack
The research was supported by the Czech
Ministry of Education, Youth and Sports within the project LC06002 and the
project of specific research of Faculty of Education, University of Hradec
Kr\'{a}lov\'{e} No. 2102/2008.
The authors are very grateful to Dita Golkov\'a and Padraig McGrath
for language corrections.

\end{document}